\documentstyle[prb,epsfig,aps,multicol]{revtex}
\begin{document}
%\preprint{SNUTP 00-XXX}
\title{Topological Quantization and Degeneracy in Josephson-Junction Arrays}
\author{M.Y. Choi}
\address{Department of Physics, Seoul National University,
Seoul 151-742, Korea}
\author{J. Choi}
\address{Department of Physics, Keimyung University,
Taegu 704-701, Korea}

\maketitle
\draft

\begin{abstract}
We consider the conductivity quantization 
in two-dimensional arrays of mesoscopic Josephson junctions, 
and examine the associated degeneracy in various regimes of the system.
The filling factor of the system may be controlled 
by the gate voltage as well as the magnetic field, and its
appropriate values for quantization is obtained by employing
the Jain hierarchy scheme both in the charge
description and in the vortex description.
The duality between the two descriptions then
suggests the possibility that the system undergoes 
a change in degeneracy
while the quantized conductivity remains fixed.
\end{abstract}
\bigskip
\thispagestyle{empty}

%{\tiny \noindent 
%74.50.+r Proximity effects, weak links, tunneling phenomena, and Josephson effects\\[-\baselineskip]
%74.25.Fy Transport properties (electric and thermal conductivity, thermoelectric effects, etc.)\\[-\baselineskip]

\begin{multicols}{2}
\narrowtext

During past decades two-dimensional arrays of superconducting grains, 
weakly coupled by Josephson junctions, have been studied extensively.\cite{proceeding} 
In particular recent advances in
fabrication techniques make it possible to control the physical 
parameters of the arrays, providing a convenient model system
for the study of charge and vortex configuration and dynamics.
When the dimensions of the superconducting grains and the capacitances
involved are small, the associated charging energy is non-negligible
and quantum dynamics comes into play at the macroscopic
level.\cite{Schonx90}  In such an array of ultra-small junctions,
frustration can be introduced not only by applying magnetic fields
but also by inducing external charges; these control the numbers of 
vortices and charges (Cooper pairs), leading to
interesting dynamic responses.\cite{fazio}
In appropriate regimes the Hall conductivity as well as the dc
component of the voltage or of the current may be quantized,
and the possibility of the corresponding quantum Hall effect \cite{qhe}
as well as the giant Shapiro steps and giant inverse Shapiro steps \cite{GSS}
has been pointed out.

This work examines such conductivity quantization
in various regimes of the system,
with regard to the associated degeneracy. 
We thus consider a Josephson-junction array with the junction
capacitance between nearest-neighboring grains 
dominant over the self-capacitance of each grain;
the system is characterized by the charge-vortex duality,
which is manifested by transforming charge variables 
into vortex variables.
Here in the presence of
both the charging energy and the Josephson coupling energy,
charges or vortices may form incompressible quantum liquids
and display quantization of the Hall conductivity
at appropriate values of the filling factor,
which may be controlled 
by the gate voltage as well as the magnetic field.\cite{limit}
We employ the Jain hierarchy scheme\cite{jain} to obtain the quantization
values of the filling factor both in the charge
description and in the vortex description.
The duality between the two descriptions then
suggests that the system may undergo a change in degeneracy
while the quantized conductivity remains fixed.

We begin with an $L\times L$ square array ($L^2 \equiv N$) 
with the Josephson coupling $E_J$
and the charging energy $E_C \equiv e^2/2C$, 
in the limit that the self-capacitance $C_0$
is negligibly small compared with the junction capacitance $C$.  The 
system in the uniform transverse magnetic field 
${\bf B}\equiv {\bf \nabla} \times {\bf A}$ 
is described by the Hamiltonian
\begin{eqnarray} 
H = & & 4E_C \sum_{{\bf r},{\bf r}'} (n_{\bf r} - f_c)G_{{\bf r},{\bf r}'}
                               (n_{{\bf r}'} - f_c) \nonumber\\
   & & - E_J \sum_{\langle {\bf r},{\bf r}'\rangle} 
                  \cos(\phi_{\bf r}-\phi_{{\bf r}'}-A_{{\bf r},{\bf r}'}),
 \label{charge}
\end{eqnarray}
where the (excess) Cooper pair number $n_{\bf r}$ at site ${\bf r}$
is conjugate to the phase $\phi_{\bf r}$ and $G_{{\bf r},{\bf r}'}$
is the lattice Green's function. The (uniform) charge frustration $f_c$ is
related to the externally induced charge $Q$ on each grain via 
$f_c \equiv Q/2e$, whereas the plaquette sum of the bond angle 
$A_{{\bf r},{\bf r}'} \equiv (2\pi/\Phi_0)
     \int_{\bf r}^{{\bf r}'} {\bf A}\cdot d{\bf l}$ 
gives the the 
number of the flux quantum $\Phi_0 \equiv 2\pi\hbar c/2e$ or the
(uniform) magnetic frustration $f_v$ according to 
$\sum_p A_{{\bf r},{\bf r}'}=-2\pi f_v$. 
Here the system is described in terms of
charge variables; in this description the charges (Cooper pairs)
interact via the 2D Coulomb potential of strength $E_C$ while the
Josephson coupling $E_J$ provides the kinetic energy for them.

When the charging energy is smaller than the Josephson coupling energy,
it is convenient to use the dual description by means of the
vortex Hamiltonian~\cite{dual}
\begin{eqnarray} 
H= & & 2\pi E_J \sum_{{\bf R},{\bf R}'} (n^v_{\bf R} - f_v)G_{{\bf R},{\bf R}'}
                               (n^v_{{\bf R}'} - f_v) \nonumber\\
   & & - \frac{2}{\pi^2} E_C \sum_{\langle {\bf R},{\bf R}'\rangle} 
                  \cos(\phi^v_{\bf R}-\phi^v_{{\bf R}'}-A^v_{{\bf R},{\bf R}'}),
\label{vortex}
\end{eqnarray}
where the vortices, defined on dual lattice sites ${\bf R}$, are
taken for quantum mechanical particles.  Namely, the vortex charge
$n^v_{\bf R}$ and the vortex phase $\phi^v_{\bf R}$ are conjugate to
each other.  The vortex bond angle $A^v_{{\bf R},{\bf R}'}$ has been
defined in such a way that its plaquette sum (on the dual lattice) 
gives the induced charge on the enclosed grain or charge frustration,
$\sum_p A^v_{{\bf R},{\bf R}'}=-2\pi f_c$.
The vortex bond angle may also be expressed as the line integral
of the vortex vector potential ${\bf A}_v$:
$A^v_{{\bf R},{\bf R}'} = (2\pi/2e)
     \int_{\bf R}^{{\bf R}'} {\bf A}_v\cdot d{\bf l}$,
where the corresponding vortex magnetic field 
${\bf B}_v \equiv {\bf \nabla}\times {\bf A}_v$ is simply the
induced surface charge density $\rho$, so that the flux through a grain
is simply the total induced charge on that grain,
$\int {\bf B}_v\cdot d{\bf a} = Q$.
In this vortex Hamiltonian, the roles of the charging energy
and the Josephson energy are reversed: The latter describes interactions
between vortices whereas the former provides kinetic energy.
The resulting charge-vortex duality between Eqs.~(\ref{charge})
and (\ref{vortex}) has been shown to give interesting consequences
such as universal conductivity,~\cite{SI} 
persistent current and voltage,~\cite{pc} and giant Shapiro
steps and inverse steps in response to applied ac currents.~\cite{GSS} 

In the absence of the Josephson coupling ($E_J =0$), 
the ground state of the system, determined by the first term
of the Hamiltonian in Eq. (\ref{charge}), forms an insulating 
charge lattice.
When $f_c= p/q$ with $p$ and $q$ relatively prime, the charge
lattice has the $q\times q$ structure, in general with $q$-fold degeneracy.  
On the other hand, in the opposite case that the charging energy
is absent ($E_C =0$), the ground state is determined by the second term of
Hamiltonian (\ref{vortex}), forming an $s\times s$ vortex lattice
for $f_v =r/s$ (again with relatively prime $r$ and $s$).
Thus obtained is a superconducting state with $s$-fold degeneracy.
Obviously, these degeneracy factors may also be obtained from the
dual descriptions.  For example, in the latter case of $E_C =0$,
Eq. (\ref{charge}) reduces to the Hamiltonian for a (tight-binding) 
charged particle in a magnetic field, which is well known to display 
$s$-fold degenerate energy spectra.~\cite{ost}

In the presence of both charging energy and Josephson energy,
the latter provides kinetic energy of charges or the former 
provides that of vortices, which tends to destroy the charge or
vortex lattice structure.  
Unlike in a continuum system,~\cite{conti} vortices in the array system 
considered here are generally accepted as rather well-defined 
point-like objects with finite effective mass.\cite{Eckern89,lattice}  
%It feels frictional force as well as non-dissipative
%transverse force in its motion, although there has been long-standing
%controversy as to the actual determination of the
%latter in a homogeneous system.~\cite{Thoule96,Volovik} 
In particular, vortices as well as charges have been argued to be bosons,
possessing hard cores in the appropriate regimes.~\cite{qhe}
Accordingly, when the two energies are
comparable to each other, we have strongly interacting
particles (charges or vortices), 
which have been suggested to form a quantum liquid in the ground state.
In this case it is convenient to write the Hamiltonain (\ref{charge})
or (\ref{vortex})
in the second quantized representation:
\begin{equation} \label{both}
H= t \sum_{\langle i,j\rangle} e^{-iA_{ij}} b_i^{\dag} b_j 
  + u \sum_{i,j} (n_i - f)G_{ij}(n_{j} - f),
\end{equation}
where $b^{\dag}_i$ and $b_i$ are the boson (charge or vortex) 
creation and annihilation operators at (original or dual) lattice site $i$, respectively, $n_i \equiv b_i^{\dag} b_i$ is the number operator,
$A_{ij}$ is the charge or vortex bond angle, and $f$ describes 
charge or magnetic frustration.  We here consider the case of large $u$
such that bosons are well defined with small fluctuations, possessing
hard cores ($n_i = 0, 1$). This corresponds to the regime of large $E_C$
in the charge description, where
$t$ and $u$ are proportional to $E_J$ and $E_C$, respectively;
these roles of $E_C$ and $E_J$ are reversed in the vortex description.

Equation (\ref{both}) describes a two-dimensional system of $fN$
bosons, i.e., $N_c = f_c N$ Cooper pairs or $N_v = f_v N$ vortices, 
each of charge $2e$ or $\Phi_0$,
in a uniform magnetic field ${\bf B}$ or ${\bf B}_v$.  
Here each vortex carries one magnetic flux quantum
$\Phi_0$ or each charge one vortex magnetic flux quantum 
$2e$.
Thus we have $N_c$ particles together with $N_v$ flux quanta or
vice versa, leading to the charge or vortex filling factor
$\nu_c \equiv N_c/N_v = f_c/f_v$ and
$\nu_v \equiv N_v/N_c = \nu_c^{-1}$
The usual (charge) Hall conductivity thus reads
\begin{equation}
\sigma_H = \nu_c \frac{(2e)^2}{h},
\end{equation}
whereas the vortex Hall conductivity, given by the inverse
\begin{equation} \label{recip}
\sigma^v_H = \nu_v \frac{\Phi_0^2}{h} = \frac{c^2}{\sigma_H},
\end{equation}
simply corresponds to the Hall resistivity.

Via the Jordan-Wigner transformation,
the boson system described by the Hamiltonian (\ref{both}) can be
mapped into a fermion system with an additional gauge field, which
is possible owing to the hard-core condition.~\cite{jw} 
Namely, we attach $\alpha$ flux quanta to each boson, where $\alpha$
is an odd integer, and transform the boson into a fermion. 
This results in $N_c$ fermions (charges) 
with the effective number of flux quanta given by 
$N_{\phi,eff}^c= |N_v - \alpha N_c |$
or $N_v$ fermions (vortices) with 
$N_{\phi,eff}^v=|N_c -\alpha N_v |$ flux quanta,
and such a system of interacting fermions in a magnetic field
is expected to form an incompressible quantum fluid for
an appropriate filling factor.
In the simple case the ground state is described by the
Laughlin wave function
\begin{equation}
\Psi = \prod_{i<j}^{fN} |z_i -z_j|^{\alpha} (z_i -z_j)^m
          \exp\left[-\frac{1}{4\ell^2} \sum_i^{fN} |z_i |^2 \right]
\end{equation}
where $z_i$ represents the complex coordinate
of the $i$th particle (Cooper pair or vortex) and
$\ell \equiv \sqrt{\hbar c/2eB}$ or $\sqrt{\hbar c/\Phi_0 B_v}$
is the magnetic length.  Here the odd integer $m$ is related to
the filling factor via $\nu = (m+\alpha)^{-1}$, giving the
fractional quantization of the Hall conductivity
at even-denominator filling factors.

According to the topological argument,\cite{tknn} 
fractional quantization associated with the filling factor 
$1/2k$ requires that the ground-state
wave function should be multi-valued on a torus, possessing
$2k$ components.\cite{thouless} 
Here it is of interest to note that such fractional quantization of
the vortex Hall conductivity
corresponds to the integer quantization of the charge Hall conductivity.
In particular the relation between the two filling factors,
$\nu_c = \nu_v^{-1}$, suggests the possibility of different
topological degeneracy depending on the description, in terms of either
charges or vortices.
To examine such possibility, we adopt the Jain hierarchy~\cite{jain} to obtain 
appropriate filling factors for quantization. 
With $m{-}1$ flux quanta attached to each fermion, the remaining 
flux quanta gives the net number 
$N_{\phi,net}^{c,v} = N_{\phi,eff}^{c,v}-(m-1)N_{c,v}$.
Note that attaching $m{-}1$ flux quanta to a fermion corresponds to
attaching $\alpha{+}m{-}1 \equiv (2k{-}1)$ flux quanta to the boson 
(original charge or vortex), thus transforming it into a fermion.
For charges, this leads to
$N_{\phi,net}^{c}/N = f_v - (\alpha +m-1)f_c$ for $\alpha f_c < f_v$
and $N_{\phi,net}^{c}/N = (\alpha -m+1)f_c -f_v$ for $\alpha f_c > f_v$;
for vortices, the roles of $f_c$ and $f_v$ are reversed. 
The net filling factor is then given by
\begin{equation} \label{netfill}
\nu_{c,v}^{(net)} \equiv \frac{N_{c,v}}{N_{\phi,net}^{c,v}}
  = \left\{\begin{array}{ll}
     \frac{\displaystyle \nu_{c,v}}{\displaystyle 1-(\alpha +m-1)\nu_{c,v}}, &~~ \alpha f_{c,v} < f_{v,c} \\ \\
     \frac{\displaystyle \nu_{c,v}}{\displaystyle (\alpha -m+1)\nu_{c,v}-1}, &~~ \alpha f_{c,v} > f_{v,c} ,
    \end{array} \right.
\end{equation}
where $\nu_{c,v} \equiv f_{c,v}/f_{v,c}$ is the bare filling factor
for charges or vortices.
To obtain quantization, we should have an integer number of filled
Landau levels, i.e., $\nu^{(net)} = n$. 
With this condition, Eq. (\ref{netfill}) yields the values of
the filling factors appropriate for quantization, 
which, for both charges and vortices, obtain the form
\begin{equation} \label{fill}
\nu = \frac{1}{(2k-1)\pm \frac{1}{n}}
\end{equation}
with $k = 1, 2, 3, \ldots$.
When only the lowest Landau level is filled ($n=1$), the above result
reduces to $\nu = 1/2k$, reproducing the values of the Laughlin state, 
as expected.
On the other hand, in case that all the flux quanta are attached
($n \rightarrow \infty$),
we have $\nu = 1/(2k-1)$, implying that odd-denominator values are
also possible.  

In the charge description, relevant for the case 
of sufficiently large $E_C$, 
we have fractional quantization of the
Hall conductivity at $\nu_c = (2k-1\pm 1/n)^{-1}$.  
In the opposite case that $E_J$ is sufficiently larger than $E_C$,
the vortex description is applicable, yielding quantization (of the
charge Hall conductivity) at $\nu_c = (2k-1\pm 1/n)$.  In this
regime, however, the topological character is determined by the
vortex configuration, and may not be the same as that
appearing by the charge configuration.
As an example, we consider the case $f_c = 1/2$ and $f_v = 1/3$,
leading to the filling factors $\nu_c = 3/2$ and $\nu_v =2/3$.
In the charge description, Eq. (\ref{fill}) with $k=1$ and $n=3$
gives $\nu_c =3/2$, implying (charge) Hall conductivity quantization 
at this value, with two-fold topological degeneracy. 
Similarly, the value $\nu_v =2/3$
can be obtained from Eq. (\ref{fill}) with $k=1$ and $n=2$,
thus suggesting quantization of the vortex Hall conductivity at
$\nu_v=2/3$, with three-fold degeneracy. 
Via Eq. (\ref{recip}), this corresponds to quantization of
the charge Hall conductivity at $\nu_c =3/2$.
Accordingly, the quantization (of the charge
Hall conductivity) remains the same in the two regimes, but the 
topological degeneracy is apparently different.

Generalizing the above, we consider a system with given values of 
$f_c$ and $f_v$, such that the quantized filling factors are given by 
$\nu_c = f_c/f_v= p/q$ and $\nu_v =\nu_c^{-1}=q/p$ 
with $p$ and $q$ relatively prime.
When $E_J/E_C$ is sufficiently small, the charge description is
appropriate and gives $q$-fold degeneracy.  
For sufficiently large $E_J/E_C$, on the other hand, we have $p$-fold
degeneracy from the vortex description.
This is reminiscent of the problem of a charged particle in
a periodic (lattice) potential under a magnetic field:~\cite{harper} 
In the tight-binding limit, where the potential is sufficiently
strong compared with the kinetic energy, the system is described
by Harper's equation with the frustration parameter $f$,
displaying $s$-fold degeneracy. 
(Here $f=r/s$ is again the flux per plaquette in units of the flux quantum.)
In the opposite limit of weak potential, the system is
still described by Harper's equation, manifesting duality,
but $f$ is replaced by $f^{-1}$, leading to $r$-fold degeneracy.
Since the Josephson coupling corresponds to the kinetic energy
of charges, the duality between the two regimes in the array
is indeed analogous to that in Harper's equation.
In the regime $E_J/E_C \ll 1$,
the ground state of the system should be an insulator; for 
$E_J/E_C \gg 1$ it should be a superconductor.
The system is thus expected to 
undergo a phase transition between the insulating state and 
the superconducting one as $E_J/E_C$ is varied, suggesting 
the duality present between the two states.  
Assuming a single transition, we expect the system to be
self-dual at the
critical value $(E_J/E_C)_c$, which depends upon the frustration 
parameters $f_c$ and $f_v$.
It then follows that the system with quantization at 
$\nu_c =p/q$ possesses $q$-fold degeneracy for $E_J/E_C < (E_J/E_C)_c$
and $p$-fold one for $E_J/E_C > (E_J/E_C)_c$.

In summary, we have considered the possibility of conductivity quantization
in a two-dimensional array of Josephson junctions, 
and examined the associated degeneracy in various regimes of the system.
In the presence of
both the charging energy and the Josephson coupling energy,
charges or vortices
%, transformed into fermions via the Jordan-Wigner transformation, 
may form incompressible quantum liquids
and display quantization of the Hall conductivity
at appropriate values of the filling factor,
which may be controlled 
by the gate voltage as well as the magnetic field.
Adopting the Jain hierarchy scheme, we have obtained the quantization
values of the filling factor both in the charge
description and in the vortex description. 
The duality between the two descriptions has then been shown to
suggest the interesting possibility that the degeneracy of the system
can change while the quantized conductivity remains fixed.

%\acknowledgements
MYC thanks D.J. Thouless
for the hospitality during his stay at University of Washington,
where this work was performed.
JC thanks D. Belitz for the hospitality during his stay at University
of Oregon.
This work was supported in part by
the Ministry of Education through the BK21 Program
and by the National Science Foundation Grant DMR-9815932.

%===========================================================================

%\begin{center}

%\begin{figure}
%\vspace{0.5cm}
%\centerline{\epsfig{width=10cm,file=fig2a.eps}}
%\centerline{\large (a)}
%\vspace{1cm}
%\centerline{\epsfig{width=10cm,file=fig2b.eps}}
%\vspace{-0.3cm}
%\centerline{\large (b)}
%\vspace{0.5cm}
%\caption
%{$IV$ characteristics of $N{\times}N$ square Josephson-junction arrays
%Lines are merely guides to the eye.}
%\label{fig:IV}
%\end{figure}

%\end{center}
\end{multicols}


\begin{references}
\bibitem{limit}
The limitations and experimental observability have been discussed
in Ref.~\onlinecite{qhe}.  Here we assume the validity of this
proposal, and examine the consequence
of applying the Jain hierarchy.
\end{references}
\end{document}